\documentstyle[prl,aps,twocolumn,epsf,floats]{revtex}
\newcommand{\beq}{\begin{equation}}
\newcommand{\eeq}{\end{equation}}
\newcommand{\bea}{\begin{eqnarray}}
\newcommand{\eea}{\end{eqnarray}}

\newcommand{\etal}{{\em et al.}}

\def\tit#1#2#3#4#5{{#1} {\bf #2}, #3 (#4)}

\def\prl{Phys.\ Rev.\ Lett.\ }

\def\prb{Phys.\ Rev.\ B}

\def\fracc#1#2{{#1}/{#2}}

\begin{document}
\draft

\twocolumn[\hsize\textwidth\columnwidth\hsize\csname @twocolumnfalse\endcsname

\title{Two-dimensional periodic frustrated Ising models in a
transverse field}
 
\author{R. Moessner,$^1$ S. L. Sondhi$^1$\ and P. Chandra$^2$}

\address{$^1$Department of Physics, Jadwin Hall, Princeton University,
Princeton, NJ 08544, USA\\ 
$^2$NEC Research Institute, 4 Independence
Way, Princeton NJ 08540, USA} 
\date{to appear in \prl}

\maketitle

\begin{abstract}
We investigate the interplay of classical degeneracy and quantum
dynamics in a range of
%effect of introducing quantum dynamics via a
periodic frustrated transverse field Ising systems at zero
temperature. We find that such dynamics can lead to unusual ordered
phases and phase transitions, or to a quantum spin liquid (cooperative
paramagnetic) phase as in the triangular and kagome lattice
antiferromagnets, respectively.  For the latter, we further predict
passage to a bond-ordered phase followed by a critical phase as the
field is tilted. These systems also provide exact realizations of
quantum dimer models introduced in studies of high temperature
superconductivity.
\end{abstract}


]

The introduction of quantum fluctuations to a frustrated system with a large
ground-state degeneracy can lead to a multitude of new quantum phases
and transitions. This occurs because the set of
degenerate states provides no energy scale and hence all perturbations
are {\em singular}. Any linear combination of the classically
degenerate states is a candidate quantum ground state.
%The spectacularly rich phase diagram of the
%quantum Hall problem provides 
The most celebrated instance of such behaviour is the quantum Hall
problem, where the degeneracy induced by a frustrated kinetic energy
leads to a plethora of strongly correlated states when lifted by
interactions and disorder.

%The study of Ising models in magnetic fields transverse to their
%couplings has played a role in deepening our understanding of quantum
%statistical mechanics reminiscent of that played by pure Ising models
%%in the context of classical statistical mechanics. This is not
%surprising: not only do they contain minimal ingredients required for
%a non-trivial quantum dynamics; in their imaginary time
%representations, they {\it are} pure Ising models, albeit in a time
%continuum limit. Their study has led to substantial progress on
%problems such as the ferro- to paramagnetic quantum phase transition
%(both in theory\cite{Sachdev99} and in experiment \cite{Aeppli98}), also
%in the presence of randomness \cite{Motrunich99} as well as in models
%with possible spin glass ordering \cite{Bhatt97}.

In this Letter we report results from the first systematic study of a
series of simple spin models which have the potential for such
behaviour. These are geometrically frustrated transverse field Ising
models. The frustrated nature of the exchanges leads to an extensive
zero-point entropy, ${\cal{S}}$, in the classical (zero-field) case
(for a review, see Ref.~\onlinecite{Liebmann86}), and they are not
ordered even at $T=0$. The quantum dynamics is provided by the
application of a magnetic field transverse to the couplings.

These models are interesting as they are the simplest possible
frustrated quantum magnets, containing the minimal ingredients
required for a non-trivial quantum dynamics; this simplicity allows
considerable progress towards their solution. Further, in addition to
the encouraging likelihood of direct experimental realization in
highly anisotropic magnetic systems, we expect that they will serve as
effective theories in other frustrated systems where a local Ising
degree of freedom can be identified. Indeed, there is intense current
interest in these models in the context of {\it Heisenberg} systems:
the square lattice fully-frustrated Ising model arises in a group of
theories of the cuprate superconductors \cite{SV/SF}.

%They are the
%simplest possible frustrated quantum magnets, as they contain the minimal
%ingredients required for a non-trivial quantum dynamics. 
%; in their
%imaginary time representations, they {\it are} pure classical Ising
%models, albeit in a time continuum limit, discussed below. 
%Their study has led to substantial progress on
%problems such as the ferro- to paramagnetic quantum phase transition
%(both in theory\cite{Sachdev99} and in experiment \cite{Aeppli98}), also
%in the presence of randomness \cite{Motrunich99} as well as in models
%with possible spin glass ordering \cite{Bhatt97}.
%In addition, there is intense current interest in these models: the
%square lattice fully-frustrated Ising model arises in a group of
%theories currently advanced in the context of high-T$_c$\
%superconductivity \cite{SV/SF}.

Specifically, we study Hamiltonians of the form \cite{Chakrabati96}:
\beq 
H = \sum_{\langle ij \rangle} J_{ij} S^z_i S^z_j + \Gamma
\sum_{i} S^x_i + h \sum_{i} S^z_i \ ,
\label{eq:hamil}
\eeq
where the $J_{ij}$ are the nearest-neighbor exchange couplings
with $|J_{ij}| = J$ and ${\prod_{\rm plaquette} (\fracc{J_{ij}}{J}}) = -1$,
$\Gamma$ is the strength of the transverse field, the $S^a$ are the 
Pauli spin operators and $h$ is the strength of a (classical) 
longitudinal, Ising symmetry breaking field. 

%In the models of interest to us, the frustrated nature of the
%exchanges leads to an extensive zero-point entropy in the classical,
%symmetric case, $\Gamma=h=0$ (for a review, see
%Ref.~\onlinecite{Liebmann86}). Any quantum dynamics then enters as a
%{\sl singular} perturbation in this highly degenerate manifold and
%can be expected to lead to a variety of new quantum phases and
%transitions between them, much as in the instance of the
%quantum Hall problem. There the degeneracy induced by a frustrated
%kinetic energy leads to a plethora of strongly correlated states when
%lifted by interactions and disorder. 

Counterintuitively, but commonly, there is an Ising version of ``order
by disorder'' {\it \`a la} Villain \cite{Villain80}, in which
configurations with long-range order allow the softest fluctuations
and hence minimize the entropy/zero point energy; in the Ising case it
is fluctuations {\it into} the ground state manifold that matter.  In
the complementary ``disorder by disorder'' scenario,
suggested by Fazekas and Anderson \cite{Fazekas74}, a disordered
classical manifold continues into a disordered quantum phase. To our
knowledge, no such disorder-free Ising spin liquid was known to exist
previous to the work described here.

To make progress in this strongly interacting problem, we have
developed or adapted a number of approaches -- variational, weak- and
strong-coupling ones as well as mappings to other models.
From our systematic study, we present three models realising different
connections between classical and quantum ordering and, among others, 
unconventional critical and
spin liquid  phases. 
%different types of behaviour by discussing a representative model for
%each. 
A more complete account of this study, including a wider range
of lattices and finite-temperature properties, will follow this Letter
\cite{Moessner99}.

Among our results, we derive a sufficient condition for quantum
ordering, which is fulfilled by a class of models, critical at $T=0$,
including the triangular lattice Ising (IAFM) antiferromagnet and the
square fully frustrated Ising magnet (FFIM).
%In this paper we describe results on three systems that realize
%both possibilities and illuminate the connection between classical
%and quantum correlations. The triangular Ising antiferromagnet (IAFM)
%\cite{Collins97} has algebraic spin correlations at $T=0$\ and
%develops order with a $\sqrt{3} \times \sqrt{3}$ unit cell at
%small $\Gamma$ illustrating a reasonable connection between classical
%criticality and quantum order; these results include earlier ones derived
%in a different context. 
The kagome IAFM and the hexagonal fully frustrated Ising magnets, both
exhibiting exponential correlations in the ground state manifold
\cite{Liebmann86}, do not fulfill this criterion and yet lead to
different quantum states: the latter orders thus ruling out a strict
connection between classical and quantum disorder. By contrast, the
former is a rare example of a {\em quantum spin liquid} in $d=2+1$, as
it remains disordered at all couplings. In the kagome problem there is
a particularly interesting twist in which a purely longitudinal field,
$0<|h|<4J$, can be used to produce a critical classical finite-entropy
phase which then orders in a further transverse field. Interesting
byproducts of the analyses include unusual phase transitions, such as
an $O(4)$ transition for the hexagonal FFIM.
% and an equivalence of the
%small $\Gamma$ triangular/hexagonal problems with quantum dimer models
%on dual hexagonal/triangular lattices.
%a similar connection on the
%self-dual square lattice, arrived at via a ``detour'' to large N
%antiferromagnets, has recently attracted interest in the context of
%high temperature superconductivity \cite{SV/SF}.

\begin{figure}
\epsfxsize=3.0in
\centerline{\epsffile{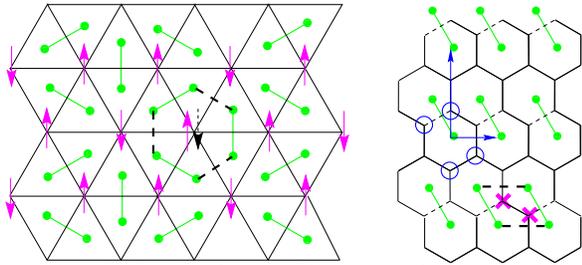}}
\caption{ Left: Maximally flippable spin configuration of the
triangular IAFM.  Dimers mark the frustrated bonds, covering the dual
lattice.  Flipping the spin in the centre, which experiences zero net
exchange field, yields the indicated dimer plaquette move. Right: The
FFIM on the hexagonal lattice, which can be thought of as a
rectangular lattice (with vectors indicated by arrows) with a basis of
four (encircled) sites. Broken lines represent antiferromagnetic
bonds, solid lines ferromagnetic ones. The dimer configuration for the
maximally flippable state with all spins aligned is shown. The
elementary dimer move corresponds to flipping the two spins indicated
by crosses.}
\label{fig:fliptrispinfin}
\label{fig:ffhexxy}
\end{figure}

\noindent
{\bf General considerations:} 
Before discussing specific lattices, we
sketch three general approaches. At small $\Gamma$, there is an Ising
analog of the saddle-point method to identify instances of order by
disorder.  In this limit we need to diagonalize the transverse field
term within the subspace of the classical ground states.  Note that
this will lead to a discontinuous change in the ground-state
properties, even though the energy levels are set by $\Gamma$, and
hence evolve continuously.  As the quantum ground state is not
perturbatively constructible, it is useful to think variationally
about different configurations. Evidently, spins able to orient in the
$x$-direction can gain energy from the transverse field. Such spins
are ``flippable'' in the $z-$representation, being connected to their
neighbors by equal numbers of satisfied and unsatisfied bonds (see
Fig.~\ref{fig:fliptrispinfin}). An intuitive variational candidate for
the ground state is thus constructed from the classical ground states
which maximize the number of flippable spins; the privileging of these
configurations, which are typically more regular, is the quantum
mechanism of order by disorder. While this kind of semiclassical
argument is suggestive, it leaves open the possibility that the true
wavefunction has the bulk of its support elsewhere in configuration
space. Note that in either case there are fluctuations in the ground
state. 

A very useful feature of the transverse field problem is that at {\it
large} $\Gamma$, there is a unique paramagnetic ground state
(polarized along $x$) separated by a gap $\Delta=\Gamma$\ from the
lowest excited states, independent of wavevector. This degeneracy is
lifted in an expansion in $J/\Gamma$: to first order, there appears a
dispersion $\epsilon({\bf k}) \sim \Delta + \sum_{\bf j} e^{i {\bf k
\cdot j}} J_{\bf 0j}$, which softens with reduced coupling. Even
without high-order computations, one can identify soft
wavevectors. Following Ref.~\cite{Blankschtein84}, these can be used
together with lattice symmetry considerations to guess at the
Landau-Ginzburg-Wilson (LGW) action governing the transition and the
symmetry-breaking pattern of the putative weak-coupling ordered phase.
This possibility does {\it not} always exist; e.g.\ an XY magnet is a
non-trivial problem on its own, which frustrates attempts to perturb
in the Ising exchange.

Finally, we derive an ordering criterion based on the mapping of the
$d$-dimensional transverse field Ising model onto a ferromagnetically
stacked model in $d+1$. This follows from the Euclidean path integral
representation, $Z = {\rm Tr}e^{-H}$, of the problem governed by the
3D classical Hamiltonian, $ H = \sum_{\langle ij \rangle, n} K^s_{ij}
S^z_i (n a_\tau) S^z_j (n a_\tau) + \sum_{i,n} K_\tau S^z_i (n a_\tau)
S^z_i (( n+1) a_\tau)$, where $a_\tau$ is the imaginary time step
introduced to obtain a discrete representation, $K^s_{ij} \propto
J_{ij}$ and the $K_\tau > 0$ are ferromagnetic. Time continuum quantum
evolution corresponds to the double limit, $a_\tau \rightarrow 0$,
$K_\tau \rightarrow \infty$ with $a_\tau e^{K_\tau} /2 = \Gamma$ held
fixed 
%and $K^s_{ij} =\infty$\ enforcing the ground-state constraint.
%\sim J_{ij} a_\tau \rightarrow 0$. 
For {\em small} $K_\tau$, the specific free energy, $F$, of this model
is given by $
%\bea
 -\beta F= {\cal S} +
{K_\tau^2}/{2} \sum_i \langle S_i(0) S_j (0) \rangle^2 +
O({K_\tau}^4),
$
with  $|K^s| =\infty$\ enforcing the ground-state constraint.
%\eea

  If the integral $I = \int d^2 r \langle S_0 S_r
\rangle^2$ diverges, the free energy above is {\sl non-analytic} as
$K_\tau \rightarrow 0$, implying that the quantum ($K_{\tau}
\rightarrow \infty$) and classical points
($K_\tau =0$) are in different phases. 
For classically disordered antiferromagnets with
exponentially decaying spin correlations at $T=0$\ even though no
single term in the series expansion becomes unbounded, the whole
series may diverge when $K_\tau$\ is beyond the radius of convergence,
and hence both ordered and disordered quantum states are allowed. In
the following we demonstrate that all three possiblities are realised
in practice by presenting an example for each one.

\noindent
{\bf Triangular IAFM: Order from criticality.} Because this lattice is
bond-sharing, the ground states of the triangular IAFM can be mapped
onto hard-core dimer coverings of its dual hexagonal lattice
(Fig.~\ref{fig:fliptrispinfin}); furthermore, since the latter is
bipartite, the ground-state manifold of this spin model admits a
height representation, which, under reasonable assumptions, implies
that the classical correlations are critical \cite{Nienhuis84}. We
have a diverging $I_{tri} = \int d^2 r/r$\cite{Liebmann86}. Our above
argument thence implies a transition separating the classical
and the quantum models.

More directly, we study the effect of small $\Gamma$ ($h=0$) for this
problem by noting that, in dimer language, states connected by the
action of the resulting Hamiltonian (Eq.~\ref{eq:hamil}) differ by a
$60^\circ$\ rotation of a triplet of dimers
(Fig.~\ref{fig:fliptrispinfin}). The degenerate problem is {\it
exactly} the hexagonal-lattice version of the Rokshar-Kivelson quantum
dimer model\cite{Rokhsar88} \bea H_{ QDM} = -t\left( |
\setlength{\unitlength}{3158sp}%
\begingroup\makeatletter\ifx\SetFigFont\undefined%
\gdef\SetFigFont#1#2#3#4#5{%
  \reset@font\fontsize{#1}{#2pt}%
  \fontfamily{#3}\fontseries{#4}\fontshape{#5}%
  \selectfont}%
\fi\endgroup%
\begin{picture}(194,185)(318,260)
\thicklines \put(468,421){\circle{18}}
\put(358,420){\line( 1, 0){108}}
\put(342,377){\circle{18}} \put(360,421){\circle{18}}
\put(396,284){\circle{18}}
\multiput(434,282)(5.48824,9.14707){11}{\makebox(8.3333,12.5000){\SetFigFont{7}{8.4}{\rmdefault}{\mddefault}{\updefault}.}}
\multiput(340,378)(5.57647,-9.29412){11}{\makebox(8.3333,12.5000){\SetFigFont{7}{8.4}{\rmdefault}{\mddefault}{\updefault}.}}
\put(434,284){\circle{18}}
\put(488,377){\circle{18}}
\end{picture}
 \rangle
\langle
\setlength{\unitlength}{3158sp}%
\begingroup\makeatletter\ifx\SetFigFont\undefined%
\gdef\SetFigFont#1#2#3#4#5{%
  \reset@font\fontsize{#1}{#2pt}%
  \fontfamily{#3}\fontseries{#4}\fontshape{#5}%
  \selectfont}%
\fi\endgroup%
\begin{picture}(194,185)(318,93)
\thicklines \put(468,117){\circle{18}}
\put(358,118){\line( 1, 0){108}}
\put(342,161){\circle{18}} \put(360,117){\circle{18}}
\put(396,254){\circle{18}}
\multiput(434,256)(5.48824,-9.14707){11}{\makebox(8.3333,12.5000){\SetFigFont{7}{8.4}{\rmdefault}{\mddefault}{\updefault}.}}
\multiput(340,160)(5.57647,9.29412){11}{\makebox(8.3333,12.5000){\SetFigFont{7}{8.4}{\rmdefault}{\mddefault}{\updefault}.}}
\put(434,254){\circle{18}}
\put(488,161){\circle{18}}
\end{picture}
|+h.c.  \right) +v\left
( |
\setlength{\unitlength}{3158sp}%
\begingroup\makeatletter\ifx\SetFigFont\undefined%
\gdef\SetFigFont#1#2#3#4#5{%
  \reset@font\fontsize{#1}{#2pt}%
  \fontfamily{#3}\fontseries{#4}\fontshape{#5}%
  \selectfont}%
\fi\endgroup%
\begin{picture}(194,185)(318,260)
\thicklines \put(468,421){\circle{18}}
\put(358,420){\line( 1, 0){108}}
\put(342,377){\circle{18}} \put(360,421){\circle{18}}
\put(396,284){\circle{18}}
\multiput(434,282)(5.48824,9.14707){11}{\makebox(8.3333,12.5000){\SetFigFont{7}{8.4}{\rmdefault}{\mddefault}{\updefault}.}}
\multiput(340,378)(5.57647,-9.29412){11}{\makebox(8.3333,12.5000){\SetFigFont{7}{8.4}{\rmdefault}{\mddefault}{\updefault}.}}
\put(434,284){\circle{18}}
\put(488,377){\circle{18}}
\end{picture}
\rangle \langle
\setlength{\unitlength}{3158sp}%
\begingroup\makeatletter\ifx\SetFigFont\undefined%
\gdef\SetFigFont#1#2#3#4#5{%
  \reset@font\fontsize{#1}{#2pt}%
  \fontfamily{#3}\fontseries{#4}\fontshape{#5}%
  \selectfont}%
\fi\endgroup%
\begin{picture}(194,185)(318,260)
\thicklines \put(468,421){\circle{18}}
\put(358,420){\line( 1, 0){108}}
\put(342,377){\circle{18}} \put(360,421){\circle{18}}
\put(396,284){\circle{18}}
\multiput(434,282)(5.48824,9.14707){11}{\makebox(8.3333,12.5000){\SetFigFont{7}{8.4}{\rmdefault}{\mddefault}{\updefault}.}}
\multiput(340,378)(5.57647,-9.29412){11}{\makebox(8.3333,12.5000){\SetFigFont{7}{8.4}{\rmdefault}{\mddefault}{\updefault}.}}
\put(434,284){\circle{18}}
\put(488,377){\circle{18}}
\end{picture}
|+
|
\setlength{\unitlength}{3158sp}%
\begingroup\makeatletter\ifx\SetFigFont\undefined%
\gdef\SetFigFont#1#2#3#4#5{%
  \reset@font\fontsize{#1}{#2pt}%
  \fontfamily{#3}\fontseries{#4}\fontshape{#5}%
  \selectfont}%
\fi\endgroup%
\begin{picture}(194,185)(318,93)
\thicklines \put(468,117){\circle{18}}
\put(358,118){\line( 1, 0){108}}
\put(342,161){\circle{18}} \put(360,117){\circle{18}}
\put(396,254){\circle{18}}
\multiput(434,256)(5.48824,-9.14707){11}{\makebox(8.3333,12.5000){\SetFigFont{7}{8.4}{\rmdefault}{\mddefault}{\updefault}.}}
\multiput(340,160)(5.57647,9.29412){11}{\makebox(8.3333,12.5000){\SetFigFont{7}{8.4}{\rmdefault}{\mddefault}{\updefault}.}}
\put(434,254){\circle{18}}
\put(488,161){\circle{18}}
\end{picture}
\rangle \langle
\setlength{\unitlength}{3158sp}%
\begingroup\makeatletter\ifx\SetFigFont\undefined%
\gdef\SetFigFont#1#2#3#4#5{%
  \reset@font\fontsize{#1}{#2pt}%
  \fontfamily{#3}\fontseries{#4}\fontshape{#5}%
  \selectfont}%
\fi\endgroup%
\begin{picture}(194,185)(318,93)
\thicklines \put(468,117){\circle{18}}
\put(358,118){\line( 1, 0){108}}
\put(342,161){\circle{18}} \put(360,117){\circle{18}}
\put(396,254){\circle{18}}
\multiput(434,256)(5.48824,-9.14707){11}{\makebox(8.3333,12.5000){\SetFigFont{7}{8.4}{\rmdefault}{\mddefault}{\updefault}.}}
\multiput(340,160)(5.57647,9.29412){11}{\makebox(8.3333,12.5000){\SetFigFont{7}{8.4}{\rmdefault}{\mddefault}{\updefault}.}}
\put(434,254){\circle{18}}
\put(488,161){\circle{18}}
\end{picture}
|\right) \nonumber \eea at the point $t=\Gamma$ and $v=0$
\cite{henabs}. The ground-state of $H_{QDM}$ at $v=t$ is known to be
an equal-amplitude superposition of all dimer coverings
\cite{fishersteph}. As the diagonal equal-time correlations at this
point are exactly those of the classical problem, the question of
fluctuation-selection reduces to determining if the point $v=0$ lies
in the same phase.  From an exact diagonalization study of clusters of
this extended problem with up to 18 spins \cite{Moessner99}, we find
that the the point $v=0$\ lies in a hexagonal dimer phase which
terminates at $v=t$. This is also the phase identified by our
heuristic semiclassical analysis (Fig.~\ref{fig:fliptrispinfin}).

Such hexagonal correlations were previously found by an LGW analysis
of ferromagnetically-stacked triangular lattices\cite{Blankschtein84},
which in the proper limit can be carried over to our model
\cite{flat}.
%equivalent to the imaginary-time path integral for the transverse field
%triangular IAFM. 
This analysis suggests that the paramagnetic phase orders via a 3D XY
transition.
%We note that a finite longitudinal field ($|h| < 6J$) induces the same
%order. 

We note that the presence of an $XY$-transition should manifest itself
in a finite-temperature critical phase where $\Gamma$ acts as a
locking field to suppress defect production \cite{Moessner99}.
Finally, we remark that these results apply, {\em mutatis mutandis},
to the square lattice FFIM.

\noindent
{\bf Kagome IAFM: Disorder from disorder.} The Kagome IAFM has
short-ranged correlations in the classical limit, with a ground state
entropy per site of $0.502k_B$, in excess of the triangular lattice
value of $0.323k_B$\ \cite{Liebmann86}. Its ground states map onto
neither a height nor a dimer model.

\begin{figure}
\epsfxsize=3in
\centerline{\epsffile{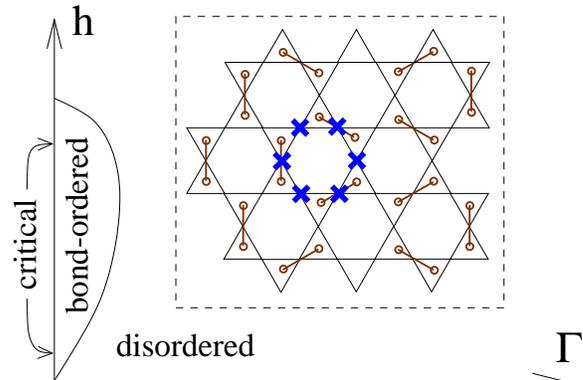}}
\caption{ Phase diagram for the kagome Ising antiferromagnet in a
field. Inset: Mapping of the kagome IAFM in a longitudinal field onto
the hexagonal lattice dimer model. 
The down spins are marked by dimers. The
non-trivial move to sixth order in perturbation theory corresponds to
flipping all the spins marked by crosses.  }
\label{fig:kagmapsmall}
\label{fig:kagphase}
\end{figure}

Interestingly, a longitudinal field $|h|<4J$ does not fully eliminate
this extensive entropy but reduces it to precisely the triangular
lattice result even as it creates a net ferromagnetic moment. Most
remarkably, it simultaneously induces critical (connected) spin
correlators. To see this, note that maximizing the polarization
requires that every elementary triangle of the lattice contain two up
and one down spin. By placing dimers on the dual lattice defined by
the triangles such that one dimer is centered on each down spin
(Fig.~\ref{fig:kagmapsmall}), one obtains dimer coverings of a
hexagonal lattice precisely as in the triangular lattice problem in
zero field, from which the above results follow.

Next, consider the effect of tilting the field somewhat, i.e.\ of
adding a small transverse component $\Gamma \ll h$. At the first five
orders in $\Gamma/h$ there is no mixing between the polarized states
and no lifting of their degeneracy. At sixth order they mix, and
the degenerate manifold Hamiltonian is precisely given by $H_{QDM}$\
with $t \propto \Gamma (\Gamma/h)^5$ and $v=0$. Hence we find that long-range
bond order of the hexagonal kind considered previously sets in
immediately.

Consider now the problem with the transverse field alone. A low-order
LGW analysis as described above fails to predict an ordering pattern:
an entire dispersionless branch of excitations \cite{reimers} goes
soft, rather than only excitations at a small number of wavevectors.
Following our previous strategy we can alternatively attempt to
identify candidate orderings by finding the maximally flippable
states. These states are precisely the `dimer states' selected by a
longitudinal field (and their Ising reversed counterparts) since only
they saturate the upper bound of two flippable spins in each triangle.
A significant restriction to this dimer manifold should manifest itself 
in correlations reminiscent of the bond-ordered phase.

To test this possibility we have carried out Monte Carlo (MC)
simulations on the Euclidean 3D path integral representation.
% $Z = {\rm
%Tr}e^{-H}$, of the problem governed by the 3D classical Hamiltonian, $
%H = \sum_{\langle ij \rangle, n} K^s_{ij} S^z_i (n a_\tau) S^z_j (n
%a_\tau) + \sum_{i,n} K^\tau S^z_i (n a_\tau) S^z_i (( n+1) a_\tau)$,
%where $a_\tau$ is the imaginary time step introduced to obtain a
%discrete representation, $K^s_{ij} \propto J_{ij}$ and the $K^\tau >
%0$ are ferromagnetic. Time continuum quantum evolution corresponds to
%the double limit, $a_\tau \rightarrow 0$, $K^\tau \rightarrow \infty$
%with $a_\tau e^{K^\tau} /2 = \Gamma$ held fixed and $K^s_{ij} \sim
%J_{ij} a_\tau \rightarrow 0$. 
As we expect any order to be strongest at small $\Gamma$, we have
primarily simulated the problem with the spatial couplings made very
large, in effect requiring that all time slices remain in the ground
state manifold.  In order to explore large $K_\tau$\ without
freezing, we use a cluster algorithm \cite{ferswen} in the time
direction so that it is the exponential growth of the correlation
length in the time direction that limits our simulations in
practice. The largest system we simulated is a stack of height 512 and
the maximal coupling was chosen such that a typical cluster size is
around 50. Each slice contains 432 spins.

\begin{figure}
\epsfxsize=3in
\centerline{\epsffile{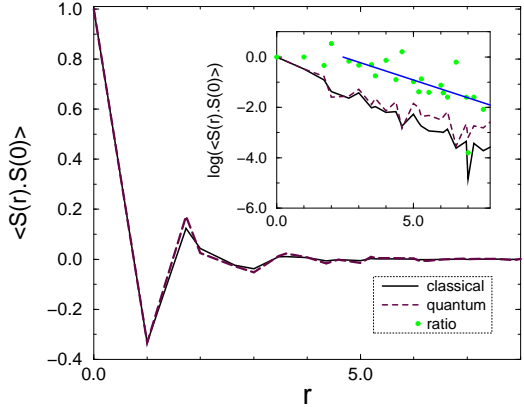}}
\caption{ Angularly averaged correlation functions (classical and
quantum) for the kagome lattice from MC simulations.  Inset: The
logarithm of the absolute value of the same functions. In addition,
the points denote the logarithm of the ratio of the classical and the
quantum correlation functions.  The straight line is a guide to the
eye.  }
\label{fig:Decaypubl}
\label{fig:Decaycomp}
\end{figure}

Fig.~\ref{fig:Decaypubl} shows the spin-spin correlation function
of this system, which is evidently tiny beyond the first few
neighbors, in stark contrast to the models known to order; e.g.\ in
the triangular case, the saturated correlation function reaches well
above 0.5 at the largest distances.

Note that the shape of the quantum correlation function mirrors that
of the classical one: their signs are the same at all points in the
region plotted. Since the classical correlations are known to decay
exponentially, and since the ratio of the two also appears to do so
but with a smaller slope (inset, Fig~\ref{fig:Decaycomp}), we conclude
that the quantum correlations, although enhanced over the classical
ones, still decay exponentially to zero with distance.

In addition, we have checked that relaxing the spatial couplings leads
to a further weakening of the correlations of the stacked model.
We have also simulated the effect of a weak longitudinal field
added to the transverse field and find that, while it induces a
magnetization per site $m \propto h$, the connected correlations
appear to remain short-ranged. At short distances, we observe
incipient correlations characteristic of the hexagonal dimer state
described above for $\Gamma \ll h$. This leads us to conjecture that,
as the field is tilted further, there is likely a continuous quantum
phase transition to the hexagonal phase.  For larger $h$,
diverging equilibration times prevent us from exploring this portion
of the phase diagram numerically.  The simplest phase diagram
incorporating all these facts within the framework discussed here is
depicted in Fig.~\ref{fig:kagphase}; evidently, this is a fruitful
topic for future work.

\noindent
{\bf Hexagonal lattice FFIM: Order from disorder.} As an example of
another classically disordered model, we consider the FFIM on the
hexagonal lattice.  This model, which maps onto a dimer but not a
height model, is obtained from a ferromagnet by changing the sign of
the exchange interaction of one bond in each hexagon, as displayed in
Fig.~\ref{fig:ffhexxy}. As the coordination of the lattice is odd, the
transverse field does not lift the degeneracy until second order in
$\Gamma$; this corresponds to flipping not an individual spin but a
pair of neighboring spins (Fig.~\ref{fig:ffhexxy}). Whereas the
entropy of the ground states is extensive, at $0.214 k_B$\ per spin
\cite{Liebmann86}, the number of maximally flippable states (for an
example, see Fig.~\ref{fig:ffhexxy}), is exponential only in the
linear size of the system.

We have carried out a LGW analysis as described above, and found that
there are four soft modes, at wavevectors $\pm(\pi/6,\pi/2),
\pm(5\pi/6,\pi/2)$, with a surprisingly large unit cell of 48
spins. The resulting LGW-Hamiltonian \onlinecite{Moessner99} is
$O(4)$-symmetric up to sixth order, where a (in $d=2+1$, dangerously
irrelevant) 48-fold symmetry breaking term appears. 
%The resulting
%ordering pattern and spin correlations do {\it not} have any obvious
%connection to the maximally flippable states.

Our MC simulations of the stacked model clearly display the pattern
predicted by the LGW analysis, indicating the presence of an ordering
transition as $\Gamma$\ is lowered \cite{lro}. The connection of this
ordering pattern with the variational analysis is rather involved
\cite{Moessner99}; here we note that the fairly non-trivial ordering
pattern is remarkable both as an instance of order by disorder and of
ordering in a quantum dimer model where the classical correlations are
short-ranged.

In summary, we have found a rich variety of behavior in frustrated
transverse field Ising models. The exciting possibility here is that
it should be possible to approximate the triangular and kagome IAFMs
fairly well not only in real, highly anisotropic magnets but also in
more general quantum frustrated systems.

We are grateful to J. Chalker for suggesting work on the kagome
magnet.  We thank D. Huse for several stimulating discussions.  We
would also like to acknowledge conversations with C.L. Henley and
S. Sachdev. RM and SLS were supported in part by grants from the
Deutsche Forschungsgemeinschaft, NSF grant No. DMR-9978074, the
A. P. Sloan Foundation and the David and Lucille Packard Foundation.


\begin{references}

\bibitem{Liebmann86}
R. Liebmann, {\sl Statistical Mechanics of Periodic Frustrated Ising
Systems} (Springer, Berlin, 1986) and references therein.

\bibitem{SV/SF} 
S. Sachdev and M. Vojta, {\sl cond-mat/9910231};
T. Senthil and M. P. A. Fisher, {\sl cond-mat/9910224}.

\bibitem{Chakrabati96}
For work on one-dimensional models, see
B.K. Chakrabati
\etal,
%, A. Dutta and P. Sen, 
{\sl Quantum
Ising Phases and Transitions in Tranverse Ising Models},
(Springer-Verlag, Berlin, 1996).

\bibitem{Villain80} 
J. Villain 
\etal,
%, R. Bidaux, J. P. Carton and R. J. Conte, 
\tit{J. Phys. -- Paris} {41}{1263}{1980}{Order as an
effect of disorder};
%\bibitem{shenderquantum}
E. F. Shender, \tit{Sov. Phys. JETP}{56}{178}{1982}{Anti-ferromagnetic
garnets with fluctuationally interacting sub-lattices}.

\bibitem{Fazekas74}
P. Fazekas and P. W. Anderson, {\sl Phil. Mag.\ }{\bf 30}, 23 (1974).

\bibitem{Moessner99}
R. Moessner, S. Sondhi and P. Chandra, {\sl in preparation}.

\bibitem{Blankschtein84}
D. Blankschtein \etal,
%, M. Ma, A.N. Berker, G.S. Grest and C.M. Soukoulis,
\prb\ {\bf 29}, 5250 (1984).

\bibitem{Nienhuis84}B. Nienhuis \etal,
%, H.J. Hilhorst and H.W.J. Bl\"{o}te,
{\sl J. Phys. A} {\bf 17}, 3559 (1984).

\bibitem{Rokhsar88}
D. S. Rokhsar and S. A. Kivelson,
\tit{\prl}{61}{2376}{1998}{SUPERCONDUCTIVITY AND THE QUANTUM HARD-CORE
DIMER GAS}.

\bibitem{henabs}
C.L. Henley in A. Gervois, M. Gingold and D. Iagolnitzer, eds.
     {\sl Book of Abstracts of STATPHYS 20 } (IUPAP Commission on
     Statistical Physics, Paris, 1998).
%This mapping was also noted independently by C. L. Henley in his talk at
%Statphys 20 (1998).

\bibitem{fishersteph}
M.E. Fisher and J. Stephenson, Phys.\ Rev.\ 132, 1411 (1963).


\bibitem{flat} Alternatively, the quantum problem can be mapped onto a
3D height model; there, only smooth surfaces are expected,
which in turn suggests the observed ordering. We are grateful to
S. Sachdev for explaining this to us.

\bibitem{reimers}
J. N. Reimers \etal,
%, A. J. Berlinsky and A.-C. Shi,
\tit{\prb}{43}{865}{1991}{Mean-field approach to magnetic ordering in
highly frustrated pyrochlores}

\bibitem{ferswen} R. H. Swendsen \etal, 
%J. S. Wang and A. M. Ferrenberg,
\tit{Top.\ Appl.\ Phys.\ }{71}{75}{1992}{NEW MONTE-CARLO METHODS FOR
IMPROVED EFFICIENCY OF COMPUTER- SIMULATIONS IN STATISTICAL-MECHANICS
}

%\bibitem{ktexp} 
%Generally, models which are critical in the classical limit appear to
%order in a transverse field. A physicist's proof can be be given by
%examining whether the discretized partition function
%$Z(K_x=\infty,K_\tau)$ posseses an expansion in powers of
%$K_\tau$. Our critical case does not, thereby signaling a phase
%transition.

\bibitem{lro} This ordering pattern is already present for only 30
stacks of 1152 spins each.  While our numerics are highly suggestive,
a definitive finite-size scaling study is out of the reach of our
current workstation due to the large size of the unit cell.

\end{references}
\end{document}